\documentclass{article}
\usepackage{graphicx} % Required for inserting images

\usepackage{amsmath}
\usepackage{amsthm}
\usepackage[colorlinks=true, allcolors=blue]{hyperref}
\usepackage[backend=biber,style=alphabetic,sorting=ynt]{biblatex}
\usepackage[margin=1in]{geometry}
\usepackage{subcaption}

\usepackage{tikz}
\usetikzlibrary{automata, positioning}

\usepackage{xcolor}

\title{Swarm Algorithms for Dynamic Task Allocation in Unknown Environments\thanks{This work was supported by the NSF under grants CCR-2139936 and CCR-2003830.}}
\author{Adithya Balachandran$^\dagger$, Noble Harasha$^\dagger$, and Nancy Lynch\thanks{Adithya Balachandran, Noble Harasha, and Nancy Lynch are with the Department of Electrical Engineering and Computer Science, Massachusetts Institute of Technology, Cambridge, MA 02139, USA {\tt\small adithyab@mit.edu, nharasha@mit.edu, lynch@csail.mit.edu}}}

\date{September 2024}

\begin{document}

\maketitle

\begin{abstract}
Robot swarms, systems of many robots that operate in a distributed fashion, have many applications in areas such as search-and-rescue, natural disaster response, and self-assembly. Several of these applications can be abstracted to the general problem of task allocation in an environment, in which robots must assign themselves to and complete tasks. While several algorithms for task allocation have been proposed, most of them assume either prior knowledge of task locations or a static set of tasks. Operating under a discrete general model where tasks dynamically appear in unknown locations, we present three new swarm algorithms for task allocation. We demonstrate that when tasks appear slowly, our variant of a distributed algorithm based on propagating task information completes tasks more efficiently than a L\'evy random walk algorithm, which is a strategy used by many organisms in nature to efficiently search an environment. We also propose a division of labor algorithm where some agents are using our algorithm based on propagating task information while the remaining agents are using the L\'evy random walk algorithm. Finally, we introduce a hybrid algorithm where each agent dynamically switches between using propagated task information and following a L\'evy random walk. We show that our division of labor and hybrid algorithms can perform better than both our algorithm based on propagated task information and the L\'evy walk algorithm, especially at low and medium task rates. When tasks appear fast, we observe the L\'evy random walk strategy performs as well or better when compared to these novel approaches. Our work demonstrates the relative performance of these algorithms on a variety of task rates and also provide insight into optimizing our algorithms based on environment parameters.
\end{abstract}

\section{Introduction}
Robot swarms, large groups of robots working together as a distributed system to reliably and efficiently accomplish tasks, have the potential to be uniquely useful in many areas such as wildfire response, land mine detection \cite{landmine}, inspection of spacecraft \cite{haghighat}, self-assembly \cite{rubenstein}, and search-and-rescue missions \cite{Couceiro_2017}. Several of these applications and others can be abstracted to the general problem of task allocation, in which abstract tasks with possibly different demands are located across an environment, and robots, which may malfunction unpredictably, must discover and assign themselves to these tasks such that the collective swarm completes all tasks efficiently. For example, in the land mine detection application, the tasks are mines to be removed or defused, while in the wildfire response application, the tasks are fires that robots need to extinguish. In several real-world scenarios, including wildfire response, these tasks may appear dynamically at unpredictable locations, and robots can move on to new tasks after finishing their current tasks. Therefore, in this paper, we investigate dynamic task allocation algorithms for robot swarms in the case where tasks appear at unknown locations dynamically. 

While several swarm algorithms for task allocation have been proposed, most of them assume either prior knowledge of task demands and locations or a fixed set of tasks at unknown locations. In \cite{nedjah}, a novel bio-inspired, distributed algorithm is described where task assignments are iteratively updated and optimized as tasks are released. However, it is assumed that task information is known to all robots instantly, which is not reasonable for applications where exploration by robots is required to discover tasks. In \cite{grace-meng,10.5555/3545946.3598680}, bio-inspired algorithms based on house-hunting and virtual pheromones are proposed for the case where static tasks are fixed at unknown locations. The latter algorithm is the task propagation (PROP) algorithm where some simple agents are evenly spaced and propagate information about tasks while other agents receive this information and choose which tasks to complete. While this algorithm was applied only to the static task case in \cite{10.5555/3545946.3598680}, it inspires our similar approach for the dynamic task setting.

In this paper, we consider a similar, but slightly more general, theoretical framework of a robot swarm as in \cite{10.5555/3545946.3598680, grace-meng} where agents, with limited communication and sensing abilities, move on a finite discrete grid and take steps in synchronized rounds. We formulate a dynamic task allocation problem where tasks can appear randomly and independently at any location and agents can assign themselves to a task at any time. Agents are free to move on to a new task at any time. The goal is to complete tasks as quickly as possible as they arrive. For this model, we introduce a modified version of the PROP algorithm from \cite{10.5555/3545946.3598680}, as well as division of labor and hybrid algorithms based on the PROP algorithm. We compare our algorithms to L\'evy random walk (RW), which is a strategy used by many organisms in nature when searching an unknown environment \cite{levy-rw}.

We propose a variant of the task propagation (PROP) algorithm from \cite{10.5555/3545946.3598680}, which includes specialized, higher-cost agents to complete tasks and simple, low-cost agents to propagate information about tasks. The specialized agents use the propagated information about tasks to choose which task to move towards probabilistically. We introduce a novel probabilistic model for specialized agents to decide when to move on to a new task, which is essential for the dynamic task setting. We also introduce a novel division of labor (DL) algorithm, where some specialized agents are allocated to perform PROP, while the remaining specialized agents perform RW. The parameter for the number of agents performing PROP is adjustable depending on the rate of tasks appearing. Finally, we also propose a hybrid algorithm, where specialized agents dynamically switch between performing PROP and RW.

We compare the PROP, DL, and hybrid algorithms to RW via simulation on a variety of task arrival rates using metrics based on how efficiently tasks are completed. We show that PROP performs significantly better than RW for low task arrival rates, but RW performs slightly better for high task arrival rates. This difference is a result of agents clumping together in the PROP algorithm, which allows agents to efficiently complete tasks when they appear slowly, but is rather inefficient when the tasks appear quickly. The DL and hybrid algorithms alleviate this issue by allocating some agents to conducting a L\'evy random walk. We show that the DL algorithm can perform better than both PROP and RW, depending on the ratio between the number of agents performing PROP and the number of agents performing RW. We also show that the hybrid algorithm where agents repeatedly use RW for a fixed amount of time before switching back to PROP can perform better than both PROP and RW. These results provide insight into the relative performance of different dynamic task allocation algorithms at varying task arrival rates. We also show how the performance of the PROP algorithm depends on the rate at which task information is propagated, with higher rates resulting in improved performance up to a point.

In Section 2, we detail related work in swarm algorithms and task allocation. In Section 3, we discuss a formal model for a robot swarm and the dynamic task allocation problem. We also introduce our algorithm based on propagation of task information and our hybrid algorithm. Section 4 contains our simulation results, which compare our algorithms to a random walk algorithm on a variety of parameters, including task arrival rate. We also include a discussion of our results in Section 4 and directions for future study in Section 5. Section 6 includes a summary of the implications of our work.

\section{Background}
Task allocation for robot swarms has been previously studied under a variety of assumptions. When information about tasks (including location) is known to all agents, some task allocation problems reduce to well-studied combinatorial optimization problems such as a set partitioning problem \cite{gerkey}. Specifically, in the dynamic task setting where task information is known, distributed algorithms have been proposed based on particle swarm optimization, where task assignments for agents are periodically updated and optimized.

Static task allocation when task information is unknown has also been studied in \cite{grace-meng, 10.5555/3545946.3598680}. In \cite{10.5555/3545946.3598680}, a house-hunting inspired task allocation algorithm is proposed, where agents explore the environment and recruit other agents to help complete tasks they find. In this approach, agents can be in 4 states, at a specified home location, exploring the environment, recruiting more agents for a task, or committed/assigned to a task. Agents transition between these states probabilistically. An algorithm based on task propagation is also discussed in 
 \cite{10.5555/3545946.3598680} where simple agents called propagators are spread out evenly in the grid to communicate information about tasks, while specialized agents use information about tasks provided by propagators to efficiently complete tasks. These specialized agents choose a task to go to depending on the distance from the agent, with closer tasks being prioritized. 

We focus our work on the case where tasks are dynamically appearing at unknown locations. The model we use is heavily inspired by \cite{grace-meng}, and our algorithms are inspired by the task propagation algorithm in \cite{10.5555/3545946.3598680}. Our algorithms allow for a computationally-efficient, distributed approach to swarm dynamic task allocation.

\section{Model and Algorithms}
We begin by discussing our discrete model for a robot swarm. Then, we will provide a formal statement of a dynamic task allocation problem and discuss a baseline random walk algorithm followed by our algorithms based on propagation of task information.

\subsection{General Model of Robot Swarm}
We work under a similar general model as in \cite{grace-meng, 10.5555/3545946.3598680}. There is a fixed finite set of agents $R$ that move on a discrete rectangular $M \times N$ grid. The vertices of the grid where agents can be located are indexed as $(x,y)$ where $0 \leq x \leq M-1$ and $0 \leq y \leq N-1$. During a transition, agents can move to an adjacent vertex or stay at the same vertex. 

Formally, we denote the set of possible states of an agent by $S_R$ and the set of possible states of a vertex by $S_V$. The \textbf{configuration} of a vertex $v$ is represented by $C(v) = (s_v, R_v, \varphi_{v})$ where $s_v \in S_V$ is the state of $v$, the set of agents at $v$ is $R_v \subseteq R$, and $\varphi_{v}: R_v \to S_R$ maps agents at $v$ to their states. 

We model a limited sensing and communication radius for agents. We assume that an agent at $(x,y)$ can obtain and see information from all vertices in $\{ (x+i, y+j) \mid -I \leq i,j \leq I \}$ where we call $I$ the \textbf{influence radius}. In particular, if $v$ is the vertex at $(x,y)$, consider a local mapping $L_v$ of vertices within the influence radius to their configurations defined by $L_v(i,j) = C(v')$ for $-I \leq i,j \leq J$ where $v'$ is located at $(x+i, y+j)$. For an agent at vertex $v$, we have the transition function $\alpha$ which maps the agent state $s_r$, the vertex $v$, and $L_v$ to a new agent state $s_r$ and a direction of motion $\{\text{Up}, \text{Down}, \text{Left}, \text{Right}, \text{Stay} \}$. Note that the transition function $\alpha$ is \textit{probabilistic} and is the same for every agent. We similarly have a transition function $\delta$ for each vertex $v$ which probabilistically maps $L_v$ into a new vertex state $s_v'$. Note that the treatment of the transition function for vertices is more general from the model in \cite{10.5555/3545946.3598680}. In \cite{10.5555/3545946.3598680}, it was assumed that vertex states were also determined by the transition function for agents $\alpha$, but here we do not impose this constraint.

In each round (which is indexed by $t$), for each vertex $v$, we apply $\delta$ to $L_v$ to determine a new vertex state as well as apply $\alpha$ to every agent to determine new agent positions and states. Note that agents and vertices transition according to $\alpha$ and $\delta$ simultaneously. We also use the directions given by $\alpha$ to update the configurations of each vertex. This process can repeat for infinitely many rounds. 

\subsection{Dynamic Task Allocation Problem Setup}

\subsubsection{Description of Tasks}
Our model of tasks is inspired by \cite{10.5555/3545946.3598680}, but we consider the dynamic task case instead of the static case. We suppose tasks are located at a subset of the grid vertices, such that there is at most one task at each vertex. Tasks will appear dynamically approximately according to an independent Poisson point process at each vertex. In particular, the number of rounds it takes for a new task to appear at each given location (either from round $0$ or after the previous task at this location disappears) is sampled independently from the exponential distribution $p(t; \lambda) = \lambda e^{-\lambda t}$ and rounded to the nearest integer. Note that this distribution has mean $\frac{1}{\lambda}$, so larger values of $\lambda$ imply that the tasks are appearing at faster rates. The \textbf{demand} $d$ of a task that appears is sampled independently from the normal distribution $\mathcal{N}(\mu_d, \sigma_d^2)$. The demand quantifies how much work by agents that the task requires. 

Once an agent is at a task, it must stay there for $t_{d}$ rounds for the amount of demand remaining for the task to decrease by $1$. As in \cite{10.5555/3545946.3598680}, we call the demand remaining on a task the \textbf{residual demand} and denote by $r$. Note that an agent must stay at a task for $t_d$ \textit{consecutive} rounds to decrease its residual demand by $1$. Once the residual demand goes to $0$, the task is completed and disappears from the grid. When an agent sees a task within its influence radius, it can detect the residual demand of the task. In our model, we encode the existence of a task at a vertex as well as its residual demand as part of the vertex's state.

\subsubsection{Performance Metrics}\label{sec:performance metrics}
We seek to dynamically distribute the agents in such a way to optimize the time required to complete tasks. We use the following two related metrics to analyze performance:
\begin{enumerate}
    \item Let $D_t$ be the sum of the residual demands of all tasks on the grid during round $t$. This is the total demand that is not satisfied by agents in this round. We consider the average total unsatisfied demand up to $T$ rounds:
\[ \mu_{\text{unsatisfied demand}, T} =  \frac{1}{T} \sum_{t=1}^T D_t. \]
This is a measure of average amount of task demand in the queue, so $\mu_{\text{unsatisfied demand}, T}$ converging to a lower value for large $T$ implies that the algorithm is more efficient at handling dynamic task demand.

    \item Suppose that $s$ tasks are completed after $T$ rounds. Let task $i$ have demand $d_i$ and be completed $t_i$ rounds after it appears for $1 \leq i \leq s$. Then, we consider 
    \[ \mu_{\text{completion time}} = \frac{1}{s} \sum_{i=1}^s \frac{t_i}{d_i} ,\]
    which is the average time to complete an individual task per unit demand. We seek to minimize $\mu_{\text{completion time}}$. Note that this is highly dependent on the parameter $t_d$.
\end{enumerate}

\subsection{L\'evy Random Walk}\label{subsec:levy}
We compare our algorithms with an implementation of the L\'evy random walk (RW) or L\'evy flight algorithm as a baseline. This is a random walk used by many organisms in nature to efficiently search an unknown environment \cite{levy-rw}, so it is quite applicable to our setup. In a L\'evy random walk, the agent takes independent steps in directions chosen uniformly at random, where the step sizes are drawn independently from a L\'evy distribution \cite{levy-rw}. This distribution has a wider tail than a Gaussian distribution, meaning that agents are more likely to take longer steps in a particular direction when compared to agents performing a Gaussian random walk. L\'evy random walk has been applied in swarm algorithms to efficiently search an unknown environment \cite{levy-swarm}.

We use a slight adaptation of the L\'evy random walk (RW) algorithm implemented in \cite{10.5555/3545946.3598680} for the dynamic task case. In particular, each agent independently perform a L\'evy walk until it sees at least one task in its influence radius. The agent then assigns itself to closest task (according to Euclidean distance) and remains there until the residual task demand for this task goes to $0$. After the task is completed, the agent resumes its L\'evy walk. 

\subsection{Task Propagation Algorithm}
We introduce a modified version of the task propagation (PROP) algorithm discussed in \cite{10.5555/3545946.3598680}. We have two types of agents, propagators and followers. As in \cite{10.5555/3545946.3598680}, propagators are simple agents who do not have the ability to complete tasks, while followers have the ability to perform tasks. We introduce propagators to model the flow of information about tasks. Propagators will use their information to guide followers to tasks. There are $MN$ propagators, one at each vertex, while there are $F$ followers.

\begin{figure}[h]
\begin{subfigure}[t]{.30\textwidth}
  \centering
  % include first image
  \includegraphics[width=.7\linewidth]{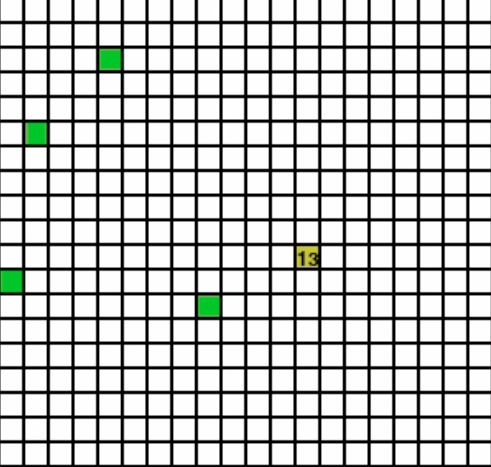}  
  \caption{A task with demand 13 appears. The propagator at this vertex receives this task information.}
  \label{fig:sub-first}
\end{subfigure}
\hspace{.03\textwidth}
\begin{subfigure}[t]{.30\textwidth}
  \centering
  % include second image
  \includegraphics[width=.7\linewidth]{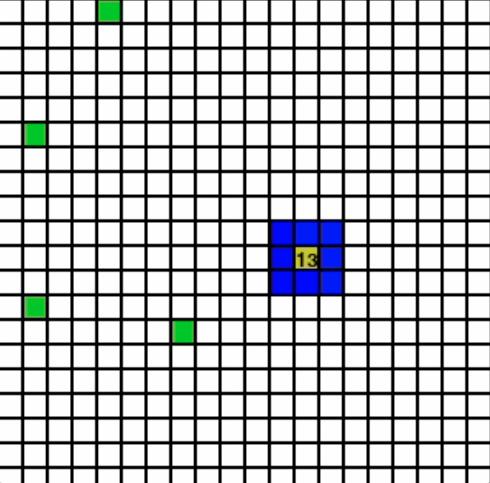}  
  \caption{After $t_p$ rounds, propagators within influence radius $I_p = 1$ receive information about this task.}
  \label{fig:sub-second}
\end{subfigure}
\hspace{.03\textwidth}
\begin{subfigure}[t]{.30\textwidth}
  \centering
  % include third image
  \includegraphics[width=.7\linewidth]{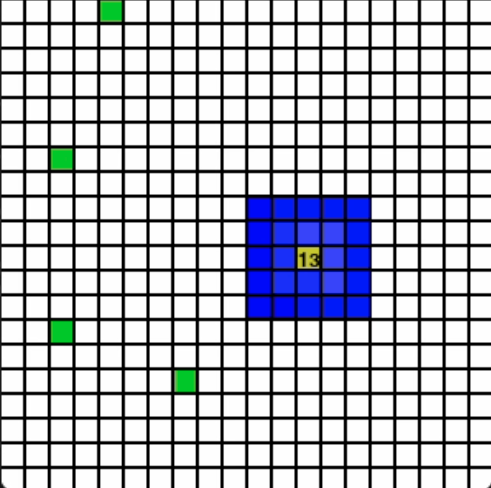}  
  \caption{After another $t_p$ rounds, propagators within influence radius $2I_p$ of the task receive task information.}
  \label{fig:sub-third}
\end{subfigure}

\begin{subfigure}[t]{.30\textwidth}
  \centering
  % include fourth image
  \includegraphics[width=.7\linewidth]{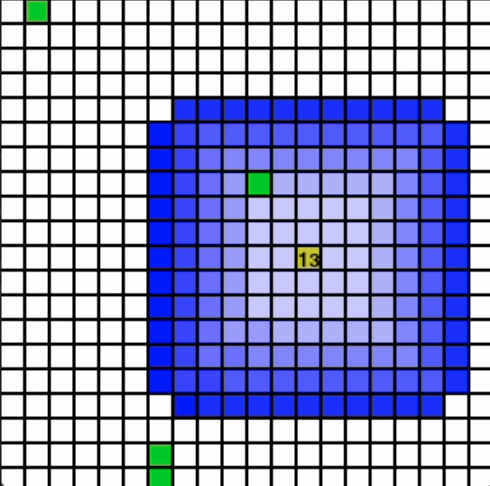}  
  \caption{A follower in the blue region moves toward the task after receiving task information from propagators.}
  \label{fig:sub-fourth}
\end{subfigure}
\hspace{.03\textwidth}
\begin{subfigure}[t]{.30\textwidth}
  \centering
  % include third image
  \includegraphics[width=.7\linewidth]{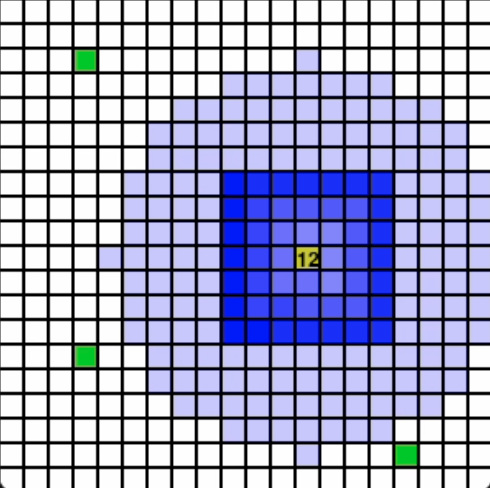}  
  \caption{The follower completes 1 unit of the task's demand, and propagators broadcast the updated residual demand.}
  \label{fig:sub-fifth}
\end{subfigure}

\caption{A simulated example of the propagator mechanism. Green squares represent followers, and yellow square represent tasks (with the number being the residual demand of the task). Blue squares are locations where the propagator agents received task information (with darker blue indicating that the task information was more recently obtained). In this example, the propagators have influence radius $I_p = 1$ and the maximum distance that task information can propagate is $d_p = 8$.}
\label{fig:prop diagram}
\end{figure}

We suppose that propagators have an influence radius $I_p$. As in \cite{10.5555/3545946.3598680}, propagators will store a mapping from task locations $(x_i, y_i)$ to their residual demands $r_i$. Every $t_p$ rounds, propagators send all new task information to all propagators within their influence radius that are at most Euclidean distance $d_p$ from the task. Note that $d_p$ is the farthest that information about a task will travel. If a propagator has the task information $(x_i, y_i, r_i)$ in its state and receives task information $(x_i, y_i, r_i')$ from another propagator, it updates the task information to $(x_i, y_i, \min(r_i, r_i'))$ (i.e., takes the minimum residual demand). This is because the residual demand of a task is nonincreasing, so the lower residual demand is newer. Once the residual demand reaches $0$, the propagator stops storing information about this task. See Figure~\ref{fig:prop diagram} for a depiction of this task propagation mechanism.

Followers can read the states of propagators to obtain information about pending tasks. Followers begin by checking if there is a task within their influence radius. If such a task exists, the follower will assign itself to the task and move towards it. If there are no such tasks, the follower will read the state of the propagator at its current location $(x,y)$. If this propagator has information about no tasks, the follower will take a step in a random direction. Otherwise, if the follower has information about tasks $(x_i, y_i)$ with residual demands $r_i$ for $1 \leq i \leq s$, then the follower will take a step toward task $(x_i, y_i)$ with probability
\[ P_{\text{move to task}}^i = \frac{ \frac{r_i}{(x-x_i)^2+(y-y_i)^2} }{ \sum_{j=1}^s \frac{r_j}{(x-x_j)^2+(y-y_j)^2}}. \]
The probability of going toward the task located at $(x_i,y_i)$ is proportional to residual demand of the task and inversely proportional to the square of the Euclidean distance to the task. This distribution is the same as used in \cite{10.5555/3545946.3598680}. It is inspired by the fact that we want agents to prioritize going to tasks that take longer to complete (i.e., tasks with larger residual demands), and we also want to prioritize tasks that are closer. Note that the agent makes independent probabilistic choices according to the above distribution in each round and only assigns itself to a task when there is a task in the agent's influence radius.

\begin{figure}[h]
  \centering
  \begin{tikzpicture}[->,>=stealth,shorten >=1pt,auto,node distance=1.5cm,
                      semithick, every node/.style={align=center}]

    \node (A) [rectangle, draw, text width = 4.5cm, fill=lightgray] {A: Move randomly};
    \node (B) [rectangle, draw, right=of A, text width = 4.5cm, fill=lightgray] {B: Move toward task according to propagator information};
    \node (C) [rectangle, draw, below=of B, text width = 4.5cm, fill=lightgray] {C: Assign to a task within influence radius};
    \node (D) [rectangle, draw, below=of A, text width = 4.5cm, fill=lightgray] {D: Complete $1$ unit of demand for task};

    \path (A) edge node {} (B)
              edge node {} (C)
              edge [loop above, looseness=7,out=60,in=120, min distance=1.5cm] node {} (A)
          (B) edge [bend right] node {} (A)
              edge node {} (C)
              edge [loop right, looseness=7,out=60,in=120] node {} (B)
          (C) edge node {} (D)
          (D) edge node {} (A)
              edge node {} (B)
              edge [bend right] node {} (C)
              edge [loop below, looseness=7,out=240,in=300,min distance=1.5cm] node {} (D); % Self-loop arrow above node A
  \end{tikzpicture}
  \caption{States of follower behavior. The follower starts in state A, and can move to state C if a task is seen within its influence radius. Otherwise, if the agent sees task information from a propagator it goes to state B. After state D, there is probability $P_{\text{stay at task}}$ of staying in state D at the same task.}
  \label{fig:follower flow diagram}
\end{figure}
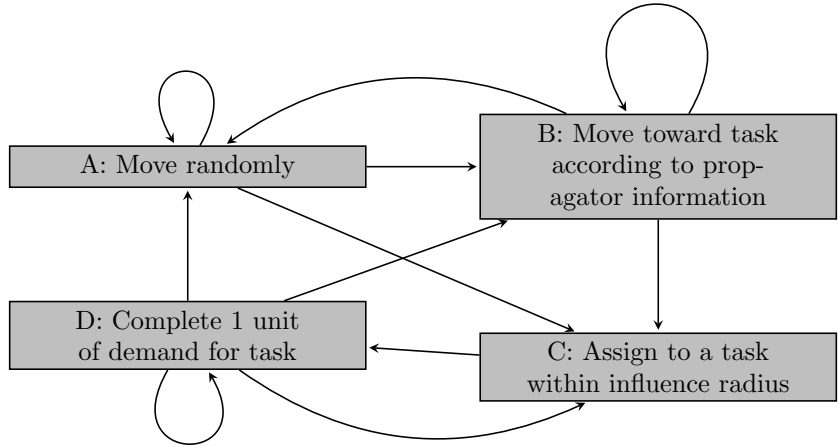

The main modification that we introduce to the PROP algorithm from \cite{10.5555/3545946.3598680} is that we allow agents to move on to a new task after completing $1$ unit of demand for the current task, which takes $t_d$ rounds. We choose to allow this instead of requiring agents to stay until the current task is completed to more efficiently allocate agent resources. The probability $P_{\text{stay at task}}$ that an agent will stay at the current task after $t_d$ timesteps have passed is given by \[ P_{\text{stay at task}}= \min \left( \frac{r_i}{k}, 1 \right) \]  where $r_i$ is once again the residual demand of the task, and there are a total of $k$ agents in the given agent's influence radius that are currently assigned to this task. The reason for this is we want ``excess'' agents for a task to move on to other tasks. If the agent decides to move on to another task, the agent's subsequent state and movement are determined by the above logic, except that this agent will never return to the task it moved on from (i.e., now $P_{\text{move to task}}^i = 0$). The different follower behaviors are illustrated as a flow diagram in Figure~\ref{fig:follower flow diagram}.

\subsection{Division of Labor Algorithm}
We introduce the Division of Labor (DL) algorithm which combines both RW and PROP. We suppose that there are $MN$ propagator agents, one at each vertex, and $F$ follower agents just like in PROP. However, we assume that a proportion $P_{\text{PROP}}$ of the follower agents are running the same PROP algorithm described above, while the remaining follower agents are conducting a L\'evy random walk (see subsection~\ref{subsec:levy}) independently of the propagators. This allows for $1 - P_{\text{PROP}}$ of the agents to be approximately evenly spaced, while the remaining agents are sensitive to communicated task information. Note that follower agents are assigned deterministically to one of the algorithms before the first round. We also observe that $P_{\text{PROP}} = 1$ is equivalent to PROP, while $P_{\text{PROP}}=0$ is equivalent to RW.

\subsection{Hybrid Algorithm}
We also introduce a hybrid algorithm between RW and PROP, where each agent switches between the PROP algorithm and a L\'evy random walk. Once again, we suppose that there are $MN$ propagator agents and $F$ follower agents. From the beginning, each agent conducts a L\'evy random walk for $t_{\text{RW}}$ rounds before switching to PROP. Additionally, every time an agent leaves a task, it follows a L\'evy random walk for $t_{\text{RW}}$ rounds before again switching to PROP. As a result of this procedure, some agents will be using a L\'evy random walk, while other agents will be using the PROP algorithm. We observe that $t_{\text{RW}} = 0$ is equivalent to PROP, and this algorithm approaches RW as $t_{\text{RW}}$ is increased.

\section{Results}
We test the performance of our algorithms using simulations and evaluate according to the performance metrics identified in subsection~\ref{sec:performance metrics}.

\subsection{Simulation Setup}
We test algorithms using simulations developed from the software in \cite{harasha_cai_sims}, which allows for simulating static task allocation algorithms for a robot swarms. Code is available at \url{https://github.com/adithyab100/swarm-dyn-task}. A visual representation of these simulations is shown in Figure~\ref{fig:prop diagram}. We tested all algorithms on a $50 \times 50$ square grid (i.e., $M = N = 50$) with $50$ agents able to perform tasks. Notably, the propagators in the PROP, DL, and hybrid algorithms are not counted among these $50$ agents since they are not able to complete tasks. The tasks have independent demands distributed according to the normal distribution $\mathcal{N}(6,3)$, and we let $t_d = 5$ so that it takes $5$ rounds for an agent to complete one unit of demand for a task. In the PROP, DL, and hybrid algorithms, we let agents have an influence radius of $2$ in all tests with an exception to the influence radius tests in subsection~\ref{subsec: comm tests}. Note that here both followers and propagators have this same influence radius, which we assumed in our testing for simplicity. Furthermore, we let $t_p = 3$ (except in subsection~\ref{subsec: comm tests}) and $d_p = 25$ so that propagation occurs every $3$ rounds and up to a maximum distance of $25$ from the task. These parameter values are similar to those used in \cite{10.5555/3545946.3598680} as they also yield computationally feasible task discovery times. 

We run each simulation for $2000$ rounds, which is enough to achieve convergence in the rtask completion time metrics (see subsection~\ref{sec:performance metrics}) for the parameter values that we test. To account for randomness, for each set of parameter values, we run $50$ tests and compute an average of the corresponding metric values. 

\subsection{Effect of Task Rate on PROP and Hybrid Performance}
To test how the rate of tasks appearing affects the performance of our algorithms, we measured the average time to complete an individual task per unit demand and the average unsatisfied task demand for $\lambda^{-1} \in \{3 \cdot 10^4, 4 \cdot 10^4, \ldots, 9 \cdot 10^4, 10^5 \}$. Recall that the tasks appear faster when $\lambda$ is larger, or equivalently when $\lambda^{-1}$ is smaller. 

\begin{figure}[h]
  \centering
\includegraphics[width=9cm]{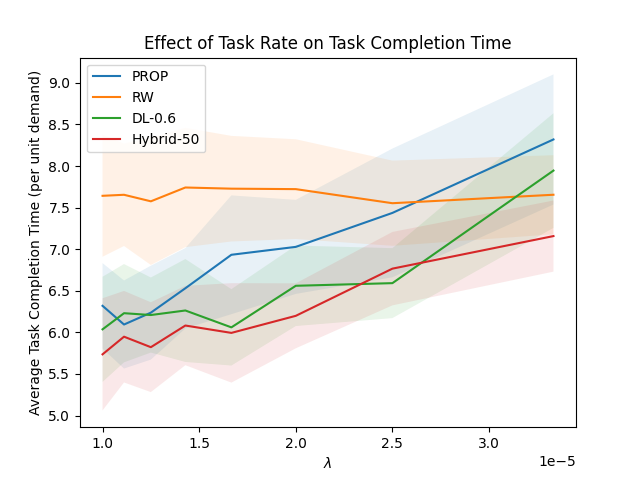}
  \caption{The effect of the rate of tasks appearing (parameterized by $\lambda$) on the average time to complete an individual task per unit demand for RW, PROP, DL (with $P_{\text{PROP}} = 0.6$), and Hybrid (with $t_{\text{RW}}=50$). We chose to highlight these specific DL and hybrid algorithms since they were overall the best performing DL and hybrid algorithms, respectively, that we tested on this range of task rates).}
  \label{fig:task_rate_completion_time}
\end{figure}

\begin{figure}[h]
  \centering
  \includegraphics[width=9cm]{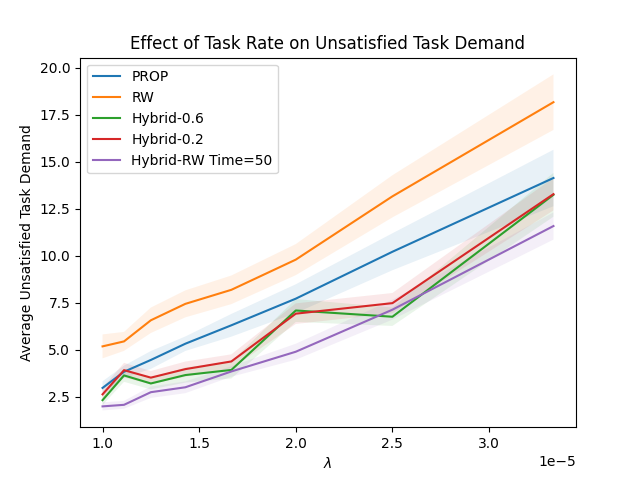}
  \caption{The effect of the rate of tasks appearing (parameterized by $\lambda$) on average unsatisfied task demand for RW, PROP, DL (with $P_{\text{PROP}} =0.6$), and Hybrid (with $t_{\text{RW}}=50$).}
  \label{fig:task_rate_unsatisfied_demand}
\end{figure}

Figure~\ref{fig:task_rate_completion_time} shows the average time to complete an individual task per unit demand for our algorithms and RW. We first observe that the PROP algorithm achieves a $17\%$ lower task completion time than RW for small $\lambda$ ($\lambda < 1.5 \cdot 10^{-5}$), which corresponds to low task rates. Near $\lambda \sim 2.5 \times 10^{-5}$, PROP and RW perform similarly, while RW has $7.9\%$ lower task completion time for even faster task rates ($\lambda \sim 3.3 \cdot 10^{-5}$). Intuitively, this occurs because when the tasks appear very slowly, PROP uses communication to relay information about each task to several agents, which ``clump up'' as they are sent to the task. RW is not able to consistently send as many agents to each task, so its performance is much worse when tasks appear slowly. However, when tasks appear faster, the clumping effect in PROP negatively affects the task completion time as no agents are sent to some tasks. RW approximately has agents evenly spaced so tasks can be reached and completed faster, on average, when they arrive fast. See subsection~\ref{subsec:discussion} for more discussion of this phenomenon.

Figure~\ref{fig:task_rate_completion_time} also shows the results of the DL algorithm where the proportion of PROP agents $P_{\text{PROP}}$ is 0.6 and the hybrid algorithm where $t_{\text{RW}} = 50$. We see that these DL and hybrid algorithms perform significantly better than both PROP and RW for task rates up to around $\lambda \sim 3.0 \times 10^{-5}$. This illustrates an interesting property that our DL and hybrid algorithms can perform better than both the PROP and RW algorithms even though all agents in both algorithms are either running PROP or RW at any time. A reason for this is that DL and hybrid both alleviate the issue we identify in PROP at high task rates by also allowing some agents to be roughly evenly distributed through random walks.

In Figure~\ref{fig:task_rate_unsatisfied_demand}, we see the performance of our algorithms on the unsatisfied task demand metric. We observe that PROP, DL, and hyrbid all have at least 21\%, 27\%, and 27\% lower unsatisfied task demand than RW, respectively, so they are are able to handle task demand more efficiently. Both DL and hybrid have very similar performance on this metric and also perform better than PROP, which is likely due to the same reasons we discussed above. Furthermore, note that while PROP becomes worse than RW for high task rates in the task completion time metric, it still performs better than RW according to the unsatisfied task demand metric. Likewise, DL and hybrid have similar performance to RW for high task rates according to the task completion time metric, but they are both significantly better than RW according to the unsatisfied task demand metric. This disparity occurs since the task completion time metric measures how fast each individual task is completed while the unsatisfied demand metric evaluates the overall completion of all tasks in the environment.

\subsection{Effect of PROP proportion on DL Performance}
To examine how the proportion of agents running the PROP algorithm affects performance in our DL algorithm, we measured our metrics when the $P_\text{PROP} \in \{0,0.1,0.2,\ldots, 1\}$ for the task rates $\lambda^{-1} \in \{3 \cdot 10^4, 5 \cdot 10^4, 9 \cdot 10^4 \}$. Note that $P_{\text{PROP}} = 0$ corresponds to RW, while $P_{\text{PROP}} = 1$ corresponds to PROP.

\begin{figure}[h]
  \centering
  \includegraphics[width=9cm]{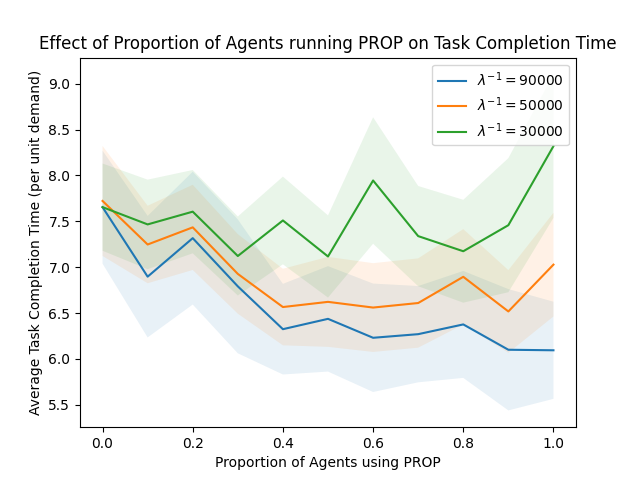}
  \caption{The effect of the proportion of agents running the PROP algorithm in our DL algorithm on the average time to complete an individual task per unit demand (for three distinct task rates).}
  \label{fig:prop_rate_completion_time}
\end{figure}

Figure~\ref{fig:prop_rate_completion_time} demonstrates the average time to complete an individual task per unit demand based on the proportion of agents using the PROP algorithm. We observe that for $\lambda^{-1} = 9 \cdot 10^4$, which is the slowest task rate, the performance of the DL algorithm according to this metric improves as the proportion of agents running PROP is increased from $0$ to $1$. This is intuitive as for extremely slow task rates, we expect that as more agents are performing PROP, the algorithm is able to more efficiently communicate information about tasks and allocate as many agents as possible to each task. However, we see that for higher tasks rates such as $\lambda^{-1} = 3 \cdot 10^4$, there is no such apparent trend. In fact, it appears that we can achieve better task completion times (than both RW and PROP) for $P_{\text{PROP}} \in [0.2, 0.9]$.

\begin{figure}[h]
  \centering
  \includegraphics[width=9cm]{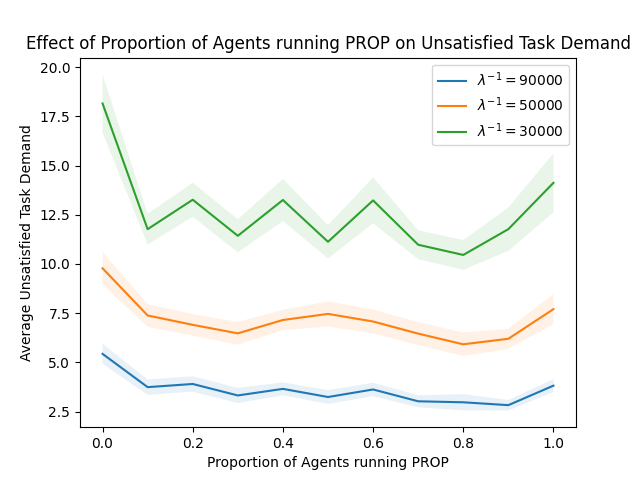}
  \caption{ The effect of the proportion of agents running the PROP algorithm in our DL algorithm on the average unsatisfied task demand (for three distinct task rates).
  }
  \label{fig:prop_rate_unsatisfied_demand}
\end{figure}

In Figure~\ref{fig:prop_rate_unsatisfied_demand}, we see graphs of the total unsatisfied task demand metric as a function of the proportion of agents using the PROP algorithm. Interestingly, for slow and medium task rates ($\lambda^{-1} \in \{5 \cdot 10^4,9 \cdot 10^4 \}$), this metric is roughly constant, while for fast task rates, there is better performance for $p_{\text{PROP}} \in [0.2, 0.9]$ similarly to the other metric. Thus, it appears that using a DL algorithm can be particularly helpful for higher task rates and does not achieve better performance than PROP at low task rates. 

\subsection{Effect of Random Walk Time on Hybrid Performance}
To examine the effect of the time spent random walking before switching to PROP in the hybrid algorithm, we simulated this algorithm for $t_{\text{RW}} \in \{5,10,25,50,100,200 \}$. Note that as $t_{\text{RW}}$ increases, at any given time, more agents will be following RW, on average. 

\begin{figure}[h]
  \centering
  \includegraphics[width=9cm]{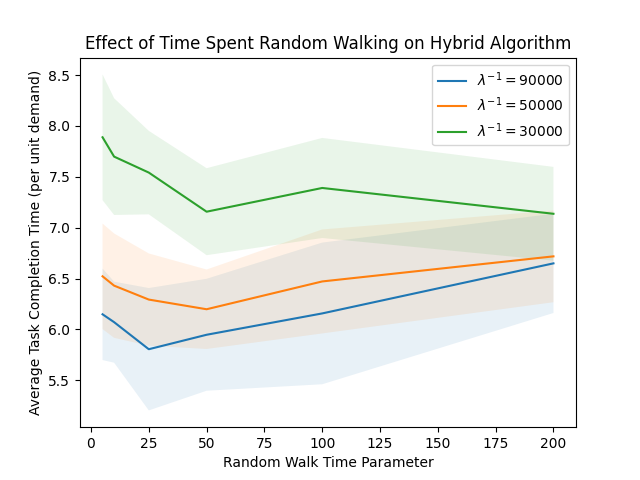}
  \caption{The effect of $t_{\text{RW}}$ in our hybrid algorithm on the average time to complete an individual task per unit demand (for three distinct task rates).}
  \label{fig:rw_time_completion_time}
\end{figure}

Figure~\ref{fig:rw_time_completion_time} demonstrates the average time to complete an individual task per unit demand based on the the parameter $t_{\text{RW}}$. We observe that for all task rates shown ($\lambda^{-1} \in \{3 \cdot 10^4, 5 \cdot 10^4, 9 \cdot 10^4 \}$), the performance of the hybrid algorithm according to this metric initially improves as $t_{\text{RW}}$ is increased from $0$. In fact, for these task rates, it appears that performance is optimized when $t_{\text{RW}} \in [25,50]$. It is intuitive that the hybrid algorithm can achieve improved performance over PROP ($t_{\text{RW}} = 0$) because this algorithm allocates some time after agents leave a task for purely exploration. This may result in agents discovering tasks faster, which can result in the PROP component of the hybrid algorithm more efficiently sending agents to task locations. 

\begin{figure}[h]
  \centering
  \includegraphics[width=9cm]{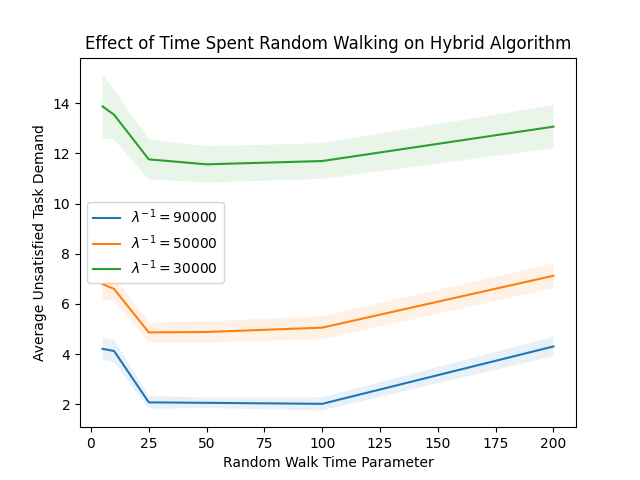}
  \caption{The effect of $t_{\text{RW}}$ in our hybrid algorithm on the average unsatisfied task demand (for three distinct task rates).
  }
  \label{fig:rw_time_unsatisfied_demand}
\end{figure}

As $t_{\text{RW}}$ becomes large enough, we see in Figure~\ref{fig:rw_time_completion_time} that the performance worsens. This is also intuitive as when $t_{\text{RW}}$ is high enough, a large proportion of the agents will be using RW at any given time, so tasks will not be completed as efficiently. We note that similar trends are also observed for the unsatisfied task demand metric in Figure~\ref{fig:rw_time_unsatisfied_demand}.

\subsection{Effect of Influence Radius on PROP Performance}\label{subsec: comm tests} 
To examine the effect of the rate of communication on performance of our algorithms, we run the PROP algorithm on a variety of influence radii values. We let $t_p = 1$, so that propagation occurs in every round. We fix the influence radius of followers to be $I = 2$ and considered all $I_p \in [1,5]$ for the task rates  $\lambda^{-1} \in \{ 5 \cdot 10^4, 9 \cdot 10^4 \}$. A greater influence radius for both propagators and followers means that agents are able to obtain task information faster and farther away from the task.

\begin{figure}[h]
  \centering
  \includegraphics[width=9cm]{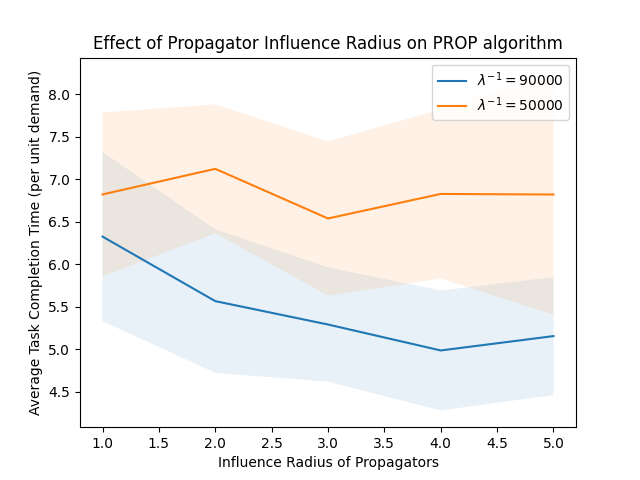}
  \caption{The effect of the propagator influence radius in the PROP algorithm on the average time to complete an individual task per unit demand (for two distinct task rates).}
  \label{fig:influence_radius_completion_time}
\end{figure}

\begin{figure}[h]
  \centering
  \includegraphics[width=9cm]{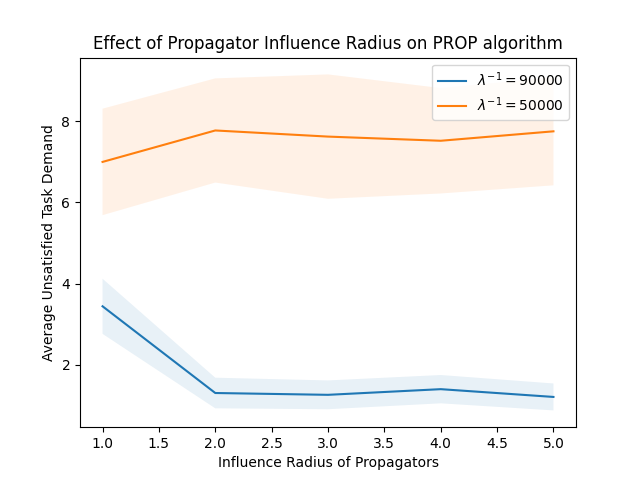}
  \caption{The effect of the propagator influence radius in the PROP algorithm on the average unsatisfied task demand (for two distinct task rates).
  }
  \label{fig:influence_radius_unsatisfied_demand}
\end{figure}

As shown in Figure~\ref{fig:influence_radius_completion_time}, when tasks appear slowly ($\lambda^{-1} = 9 \cdot 10^4$), the average time to complete an individual task per unit demand initially decreases as $I_p$ is increased from $1$. This metric appears to later stabilize, which demonstrates that when tasks appear slowly, increasing the influence radius results in improved performance in the PROP algorithm, up to a point. However, when tasks appear faster ($\lambda^{-1} = 5 \cdot 10^4$), Figure~\ref{fig:influence_radius_completion_time} demonstrates that PROP performance is unaffected by changing the influence radius according to this metric. Note that similar results for both task rates are observed in the unsatisfied task demand metric as seen in Figure~\ref{fig:influence_radius_unsatisfied_demand}. 

As expected, when tasks appear slowly, when the influence radius is initially increased, task information is propagated much faster, which allows to more efficient task completion by PROP. However, when the influence radius is further increased, it is possible that agents become overwhelmed by information about several tasks at once, so performance does not improve further. Similarly, when tasks appear faster, this effect of being overwhelmed by too much task information occurs even for smaller influence radii, which may explain why there is no improved performance seen as the influence radius is increased.

\subsection{Discussion}\label{subsec:discussion}

Our simulations demonstrate how the relative performance of our algorithms changes as the rate of tasks appearing is varied. PROP performs better than RW when tasks appear slowly, but it does not perform as well when the tasks appear faster. This is due to clustering that occurs in the PROP algorithm. When tasks appear slowly, clustering is effective because there are enough agents to send a large group of agents to each task as it arrives. However, when tasks appear quickly, clustering can be counterproductive. Especially since tasks appear dynamically, it is possible for a new task to appear in a location after a clump of agents have already started moving away to another task, which would make it take longer for the new task to be completed. Thus, the clumping effect in PROP is useful when tasks appear slowly but ineffective when tasks appear quickly. RW, on the other hand, has the agents better distributed, so its performance can be better at high task rates. Our DL and hybrid algorithms can achieve better performance than both RW and PROP, depending on the proportion of robots using the PROP algorithm and the random walk time parameter, respectively, by combining the benefits of both algorithms. The agents in the DL and hybrid algorithms that are running RW at a given time will be roughly evenly spaced, which alleviates the issue due to clumping in PROP at faster task rates. 

Our results extend the work in \cite{10.5555/3545946.3598680} to a dynamic task setting. Specifically, we use a similar model of a robot swarm as in \cite{10.5555/3545946.3598680} and adapt the PROP algorithm in \cite{10.5555/3545946.3598680} to the dynamic task case. This model assumes that task locations and demands are unknowns (unlike the model in \cite{nedjah}), which are reasonable assumptions for applications requiring exploration by agents to discover tasks. We also assume a limited communication and sensing radius like in \cite{10.5555/3545946.3598680}, which limits our potential algorithm performance, but allows for more computationally-efficient algorithms.

\section{Future Work}
Future work could evaluate the effect of other parameters on the performance of our algorithms. For example, one could analyze the effect of swarm density, the ratio of of the number of agents to the total grid area. A greater swarm density may allow PROP to send even more agents to each task, but it could also worsen the clumping effect at high task rates. Another parameter of interest is $d_p$, the maximum distance from a task that information about it propagates. For the static task case, \cite{10.5555/3545946.3598680} showed that $d_p$ has a different effect on task completion time depending on the task density. We wish to determine if an analogous result exists when tasks appear dynamically. It is also useful to further understand the effect of influence radius and rate of communication on the performance of PROP and hybrid algorithms. One particular case that can be studied is where followers still have a limited influence radius (such as $I = 2$) while propagators can instantaneously communicate task information. In fact, this setting is equivalent to removing all propagators and having followers have an influence radius of $d_p$.

One could also analyze the number of messages sent about task information in our PROP and hybrid algorithms and explore methods to reduce this message count while maintaining performance. Another area of future work is analytical bounds on the expected values of the unsatisfied demand and task completion time metrics for our PROP and hybrid algorithms. Furthermore, one could also study this problem with additional realistic assumptions such as noisy information about tasks and possible individual robot failure, investigating in simulation if different algorithms are more or less robust to these forms of noise.

With the strong performance of our DL and hybrid algorithm, it is also natural to consider other types of algorithms that combine the high-level strategies of PROP and RW. One such alternative is a adaptation of PROP in which at each time step, there is always a nonzero probability of going in any direction, even if there is no known task in a given direction. Here, as well as in the case of the DL and Hybrid algorithms, we are leveraging some amount of uniform randomness in order for agents to better explore the environment to discover new tasks and respond to them quickly.

\section{Conclusion}
Our swarm algorithms can be applied to several dynamic task allocation applications such as search-and-rescue or the problem of detecting and extinguishing wildfires. With the PROP algorithm, we have a distributed approach that can dynamically complete tasks significantly more efficiently than a L\'evy random walk when the tasks appear slowly, which is reasonable for many of these application domains. Furthermore, through the use of our DL and hybrid algorithms, we can also achieve efficient performance in comparison to L\'evy random walk when tasks appear faster. Based on our results, we can tune the parameters of these algorithms to optimize efficiency. In a physical robot swarm, these algorithms can be easily implemented -- low-cost agents with only communication abilities can be spaced out in an unknown environment as propagators, while higher-cost, task-completing agents can run implementations of the PROP, DL, or hybrid algorithms.

\section{Acknowledgement}
Special thanks to Grace Cai for her ideas for future extensions upon our earlier work on the task allocation problem, some of which were investigated in this paper.

\printbibliography

\end{document}